\documentclass[pra,aps,showpacs,twocolumn]{revtex4}
\usepackage{epsf}

\begin{document}

\title{Dynamics of a matter-wave bright soliton in an expulsive potential}
\author{L.~D. Carr and Y. Castin\\}
\affiliation{Laboratoire Kastler Brossel, Ecole Normale
Sup\'erieure, 24 rue Lhomond, 75231 Paris CEDEX 05, France}
\date{\today}

\begin{abstract}
The stability regimes and nonlinear dynamics of bright solitons
created in a harmonic potential which is transversely attractive and
longitudinally expulsive are presented.
This choice of potential is motivated by the recent creation of a
matter-wave bright soliton from an attractive Bose-Einstein
condensate (L. Khaykovich {\it et~al.}, Science 296, 1290 (2002)).  The critical branches for collapse due to
the three-dimensional character of the gas and
explosion caused by the expulsive potential are derived based on variational studies.  Particle loss
from the soliton due to
sudden changes in the trapping potential and scattering length are
quantified.  It is shown that higher order solitons can also be
created in present experiments by an abrupt change of a factor of four
in the scattering length.  It is
demonstrated that quantum evaporation occurs by nonlinear tunneling of
particles out of the soliton, leading eventually to its explosion.  
\end{abstract}


\maketitle

\section{Introduction}
\label{sec:intro}

Solitons are a central paradigm of nonlinear physics~\cite{campbell1}.  They appear in
systems as diverse as the ocean~\cite{hasegawa1}, nonlinear fiber
optics~\cite{agrawal1}, and Bose-Einstein condensates (BEC's)~\cite{denschlag1,burger1}.  Such localized waves propagate without
spreading.  They retain their form after interactions and are robustly
stable with respect to perturbation.  In this sense they are
particle-like.  In fiber optics, solitons have become the basis of transoceanic communication
systems~\cite{haus1}.  They are called {\it bright} when they represent a
local maxima in the field, and dark or gray or kink when they represent a
minima.  {\it Matter-wave} bright
solitons arise from a balance between nonlinearity and dispersion in the Gross-Pitaevskii
equation (GPE) which describes the mean field of the BEC.  Such a macrosopic
bound state of atoms is intrinsically quantum
mechanical, in that its existence depends on quantum, rather than
classical motion of the atoms.  Matter-wave bright solitons therefore represent an
intriguing combination of nonlinear and quantum physics.

Investigations of BEC's with attractive interactions have focused on
their collapse~\cite{dalfovo1,koehler1}.
In the attractive GPE, which for the case of constant potential is also known
as the focusing nonlinear
Schr\"odinger equation, it
is well known that solutions collapse in finite time~\cite{sulem1}.  This
collapse can be prevented, below a certain critical nonlinearity, by
the addition of a spatially varying external potential.  Intensive
studies of collapse dynamics~\cite{dodd1,kagan2,sackett2} were
initially inspired by experiments on the attractive BEC~\cite{bradley1,bradley2} and by
the possibility of efficient implementation of Feshbach resonances~\cite{vogels1}.
The latter use an external
magnetic field to tune the $s$-wave scattering
length which controls the strength of the nonlinearity governing the
collapse: in particular, $^{85}$Rb has been the subject of recent
experiments~\cite{roberts1,donley1}.  In $^7$Li there exists a
Feshbach resonance around
725 gauss for the state $|F=1,m_F=1\rangle$ which is
% particularly
amenable to the study of {\it stable} phenomena in attractive BEC's: the scattering length for $B=0$ is positive, and becomes
slightly negative between 150 and 525 gauss, far from the resonance
and thus far from the region in which three-body losses are
significant.  $^7$Li was the first species used to
create an attractive BEC~\cite{bradley1,bradley2,sackett1}, and it is
$^{7}$Li that has been used to create the first matter-wave bright
solitons~\cite{carr29,strecker1}.

The essential experimental method used in Ref.~\cite{carr29} to transform a stable attractive
BEC into a matter-wave bright soliton was as follows.  A BEC with positive scattering
length was condensed into a harmonic trap.  The trap was adiabatically~\cite{band1}
deformed into a cigar-shaped geometry,
{\it i.e.}, $\omega_z \ll
\omega_{\rho}$, where $\omega_z$ and
$\omega_{\rho}$ are the longitudinal and axial oscillation frequencies
of the atoms in the trapping potential,
respectively.  The
scattering length was then tuned via a Feshbach resonance to be small
and negative.  The resulting condensate, a metastable
macroscopic quantum bound state of about 5000 particles, was projected onto an {\it expulsive}
harmonic potential in the longitudinal direction.  It was then
observed to propagate, without spreading, over a distance of 1.1
mm, in contrast to an ideal gas, which is exploded exponentially
by the potential.  Thus it was the use
of the expulsive potential which provided clear evidence of a
self-trapped, soliton-like state.

In this paper the stability and dynamics of a matter-wave bright
soliton in an expulsive potential are studied.  The 3D GPE which
describes the mean field of the BEC is written as
\begin{eqnarray}
\label{eqn:gpe3d}
&&\left \{ -\frac{\hbar^2}{2 m}\nabla^2
+g\,N\,|\psi(\vec{r},t)|^2 
+\frac{1}{2}m^2[\omega_{\rho}^2(x^2+y^2)
\right .\nonumber \\
&&\left . 
+\omega_z z^2]\right \} \psi(\vec{r},t) 
=i\hbar\frac{\partial}{\partial t}\psi(\vec{r},t) \, ,
\end{eqnarray}
where $g\equiv 4\pi\hbar^2 a/m$, $a$ is
the $s$-wave scattering length tuned by the Feshbach resonance, $m$
is the atomic mass, $N$ is the number of atoms in the condensate, and $\psi$ has been normalized to
one.  Equation~(\ref{eqn:gpe3d}), in conjunction with variational
studies, numerical
integration, reduction to an effective one-dimensional problem, and the semi-classical WKB approximation, forms the basis of the work herein.
Section~\ref{sec:stability} presents the size, energy, and critical
parameters for a stable soliton.  Section~\ref{sec:nonlinear} studies the dynamical effects
particular to an expulsive potential: the equations of motion for the
soliton parameters are
derived; weak excitations are shown to be exponentially
damped; higher order solitons created by sudden changes in the
scattering length are shown to persist in the presence of the potential; and an intriguing form of macroscopic nonlinear tunneling, which we call
{\it quantum evaporation}, is described.  The latter eventually leads
to explosion of the soliton.  A form of nonlinear tunneling similar to
this has already been experimentally observed in the context of
the relaxation of condensate spin domains~\cite{stamperkurn1}.  Finally, in
Sec.~\ref{sec:conclusions} the conclusions and experimental outlook are presented.

\section{Stability Regimes}
\label{sec:stability}
The first and most important question concerning the attractive BEC is
under what conditions it remains robustly stable.  
The GPE, in general,
has proven a sufficient theoretical tool for describing the vast
majority of BEC phenomena for $T\ll T_{{\rm BEC}}$.  However, as a
three-dimensional second-order nonlinear partial differential equation it is not
amenable to analytic methods without approximation, of which the most
commonly used for the repulsive, or positive scattering length case, is the
Thomas-Fermi approximation~\cite{dalfovo1}, which neglects the kinetic
energy term in Eq.~(\ref{eqn:gpe3d}).  In the attractive case
one must find another method, since the kinetic energy and
external potential terms must dominate the mean field in order to prevent
collapse.  The variational method
is the most immediate and useful approximation~\cite{desaix1}.  A thorough
variational study of the Lagrangian dynamics of a purely gaussian ansatz has already been made by P\'erez-Garc\'ia
{\it et al.}~\cite{perez2}.  However, this work did not focus on the
case of bright soliton solutions in an expulsive potential.  To this
end a hyperbolic secant rather than gaussian ansatz shall be utilized in the
longitudinal direction.  In limits
germane to experiments this choice leads to closed form analytic
expressions for the widths and energy of the condensate.

Three separate cases are therefore considered: general three-dimensional
axisymmetric confinement, which is the case of Ref.~\cite{carr29};
$\omega_z=0$, which allows for full analytic solutions; and finally
the quasi-1D limit, which is the case of Ref.~\cite{strecker1} and the
one likely to be relevant for future
experiments on matter-wave bright solitons.

\subsection{Case I: Three-Dimensional Confinement}
\label{ssec:3d}

The Gross-Pitaevskii energy functional is defined as
\begin{eqnarray}
E_{{\rm GP}}[\psi]&
=&\int d^3\vec{r}\, 
\left\{ \frac{\hbar^2}{2m}|\nabla\psi(\vec{r}\,)|^2
+\frac{g\,N}{2}|\psi(\vec{r}\,)|^4 \right.
\nonumber \\
&+& 
\left. \frac{1}{2}m\left [\omega_{\rho}^2 (x^2+y^2)+\omega_z^2
z^2 \right ]|\psi(\vec{r}\,)|^2 \right\}\ .
\label{eqn:energy}
\end{eqnarray}
In the case of an expulsive potential $\omega_z^2<0$.  A good stationary variational ansatz for a bright soliton solution is
\begin{equation}
\psi_a(\vec{r}\,) =
\frac{1}{\sqrt{2\pi\,l_{\rho}^2\, l_z}}
\exp\left(-\frac{x^2+y^2}{2\,l_{\rho}^2}\right)
\,{\rm sech}\left(\frac{z}{l_z}\right)\, ,
\label{eqn:ansatz3d}
\end{equation}
where $l_{\rho}$ and $l_z$ are the transverse and longitudinal widths
which form the variational parameters.  Substituting
Eq.~(\ref{eqn:ansatz3d}) into Eq.~(\ref{eqn:energy}) and rescaling all
variables by the axial oscillator frequency $\omega_{\rho}$, one finds 
\begin{equation}
\label{eqn:energy2}
\epsilon_{{\rm GP}}=
\frac{1}{2\,\gamma_{\rho}^2}
+\frac{\gamma_{\rho}^2}{2}
+\frac{1}{6\,\gamma_z^2}
+\frac{\pi^2}{24}\lambda^2\,\gamma_z^2
+\frac{\alpha}{3\gamma_{\rho}^2\,\gamma_z} \, ,
\end{equation}
where all quantities have been scaled to the transverse harmonic
oscillator frequency $\omega_{\rho}$ or length
$\sigma_{\rho}\equiv\sqrt{\hbar/(m\omega_{\rho})}\,$: $\epsilon_{{\rm GP}}\equiv E_{{\rm GP}}/(\hbar\omega_{\rho})$, $\gamma_{\rho}\equiv l_{\rho}/\sigma_{\rho}$,
$\gamma_z\equiv l_z/\sigma_{\rho}$, $\lambda\equiv
\omega_z/\omega_{\rho}$, and $\alpha\equiv
Na/\sigma_{\rho}$.  Taking the partial derivatives with respect to
$\gamma_{\rho}$ and $\gamma_z$ and assuming $\gamma_{\rho}\neq 1$, one
may simplify and obtain
the equations for the extrema points:
\begin{eqnarray}
\label{eqn:root1}
&&-\gamma_{\rho}^{18}
+4\gamma_{\rho}^{14}
-\frac{2}{3}\alpha^2(-1+\frac{\pi^2}{4\,3^4}\lambda^2\alpha^4)\gamma_{\rho}^{12}
\nonumber \\
&&
-6\gamma_{\rho}^{10}
+4\gamma_{\rho}^6
-2\alpha^2\gamma_{\rho}^4
+6\gamma_{\rho}^2
=0
\end{eqnarray}
and
\begin{equation}
\label{eqn:root2}
\gamma_z=\frac{-2\,\alpha}{3(1-\gamma_{\rho}^4)}
\, .
\end{equation}
Whether these extrema are saddle points, minima, or maxima, is
determined by the Jacobean
$J\equiv(\partial_{\gamma_{\rho}}^2\epsilon_{{\rm
GP}})(\partial_{\gamma_{z}}^2\epsilon_{{\rm
GP}})-(\partial_{\gamma_{\rho}}\partial_{\gamma_{z}}\epsilon_{{\rm GP}})^2$,
which, using Eq.~(\ref{eqn:root2}) to eliminate $\gamma_z$, yields
\begin{eqnarray}
\label{eqn:jac}
&&
J=
\gamma_{\rho}^{22}
-4\gamma_{\rho}^{18}
+\frac{1}{3}\alpha^2\gamma_{\rho}^{16}
+6\gamma_{\rho}^{14}
-\frac{8}{9}\alpha^2\gamma_{\rho}^{12}
\nonumber \\
&&
-4\gamma_{\rho}^{10}
+\frac{2}{3}\alpha^2\gamma_{\rho}^{8}
+(1+\frac{4\pi^2}{3^5}\lambda^2\alpha^4)\gamma_{\rho}^{6}
-\frac{1}{9}\alpha^2
\, .
\end{eqnarray}

If Eq.~(\ref{eqn:jac}) is negative for a given solution
$(\gamma_{\rho{\rm o}},\gamma_{z{\rm o}})$ to
Eqs.~(\ref{eqn:root1}) and~(\ref{eqn:root2}), $(\gamma_{\rho{\rm
o}},\gamma_{z{\rm o}})$ is a saddle point;
if it is positive one must further examine $\partial_{\gamma_{\rho}}^2\epsilon_{{\rm
GP}}$ or $\partial_{\gamma_z}^2\epsilon_{{\rm
GP}}$.  If these second partial derivatives are positive at $(\gamma_{\rho{\rm o}},\gamma_{z{\rm o}})$ the extremum is a minimum; if they are
negative the extremum is a maximum.  As Eqs.~(\ref{eqn:root1})
and~(\ref{eqn:jac}) are high order polynomials, it is necessary to
obtain $(\gamma_{\rho{\rm o}},\gamma_{z{\rm o}})$ numerically for
given parameters $\lambda$ and $\alpha$.  For attractive interactions, such
solutions are found to constitute a local minimum of the $\epsilon_{{\rm
GP}}$ surface for a finite range of the parameters and constitute
bright solitons, provided that the spatial extent of the condensate is
smaller than the longitudinal oscillator length.  A specific example relevant to the experiment of
Ref.~\cite{carr29} shall be presented in Sec.~\ref{ssec:project}.

%%%%%%%%%%% figure 1 %%%%%%%%%%%
%
\begin{figure}[hb]
\begin{center}
\epsfxsize=8cm \leavevmode \epsfbox{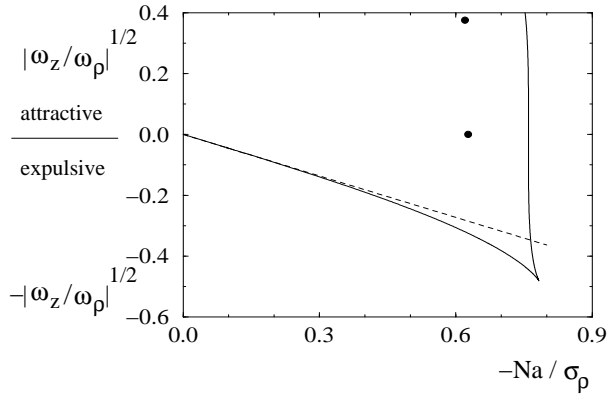}
\caption{\label{fig:crit}  Stability diagram for parameter space of
matter-wave bright soliton; the upper half
plane $y>0$ represents the case of an attractive harmonic trapping potential,
while $y<0$ represents that of an expulsive harmonic potential.  Solid line:
explosion branch (left) and collapse branch (right) obtained from
three-dimensional variational studies.  Dashed line:
quasi-one-dimensional variational prediction for explosion branch.  Filled circles:
collapse points obtained by imaginary time relaxation of the
Gross-Pitaevskii equation.  The latter differ by $\sim 17.5\%$ from
variational predictions.}
\end{center}
\end{figure}
%
%%%%%%%%%%%%%%%%%%%%%%%%%%%%%%%%%%%%%%%%

The critical points for collapse caused by the mean field, and, in the case of $\omega_z^2<0$,
for explosion due to the expulsive potential, may be found by simultaneously solving Eq.~(\ref{eqn:root1})
and Eq.~(\ref{eqn:jac}) for $J=0$.  It is expedient to use the
resultant to eliminate $\gamma_{\rho}^2$, which
yields a single implicit equation in $\lambda$ and $\alpha$.  The numerical
solution to this polynomial equation is plotted in
Fig.~\ref{fig:crit} (solid line).  The outer branch signifies collapse; the second,
inner branch, which exists only for an expulsive potential, represents explosion.  For $\omega_z^2>0$ the soliton can be made
between the vertical axis and the collapse branch; for $\omega_z^2<0$ it can
be created in the region between collapse and explosion.  The cusp represents the doubly critical point beyond which a soliton
cannot be formed, and can be calculated analytically to be
\begin{eqnarray}
\label{eqn:crit1}
&& -N a/\sigma_{\rho}
=\frac{6^{1/2}}{5^{5/8}}\sqrt{3-\sqrt{5}}
\simeq 0.7830 \, ,
\\
\label{eqn:crit2}
&&-\sqrt{\left|\frac{\omega_z}{\omega_{\rho}}\right|}
=-\sqrt{\frac{3}{2\pi}}(\sqrt{5}-2)^{1/4}
\simeq -\sqrt{\frac{1}{4.3113}}
\, .
\end{eqnarray}
It is therefore clear that there is a limited
parameter range under which bright solitons can be studied in harmonic
potentials.  This
range, as demonstrated in Refs.~\cite{carr29} and~\cite{strecker1}, is experimentally realizable.

Other features of interest not shown in Fig~\ref{fig:crit} are the critical branch for
collapse for oblate traps, $\omega_z/\omega_{\rho}\gg 1$, which approaches the
vertical axis as $\omega_z/\omega_{\rho}\rightarrow +\infty$, and the critical point for
the isotropic case $\omega_z/\omega_{\rho}=1$, which is found by the above methods to be
$-Na/\sigma_{\rho}=0.6501\cdots$.  The isotropic case has been
extensively studied, and is determined from a purely
gaussian hypothesis~\cite{baym1} to be $-Na/\sigma_{\rho}\simeq 0.671$.  However, the value found from numerical studies~\cite{ruprecht1,eleftheriou1}
of Eq.~(\ref{eqn:gpe3d}) is $\sim 0.575$.  The filled circles in
Fig.~\ref{fig:crit} mark critical points for collapse determined by
imaginary time relaxation of Eq.~(\ref{eqn:gpe3d}), and differ by
between 12.5\% (isotropic case) and 17.5\% (zero longitudinal
potential, following section) from the variational predictions.  A
more detailed description of our numerical methods is made in
Sec.~\ref{sec:nonlinear}.  Note that even lower
critical values
for collapse have been found
in experiments for which the initial state differs from the
equilibrium one~\cite{koehler1,donley1}: the variational studies presented herein make
similiar predictions, as shall be illustrated in Sec.~\ref{ssec:project}.

\subsection{Case II: Longitudinal Freedom}
\label{ssec:free}

In the case of $\omega_z=0$, Eqs.~(\ref{eqn:root1}) and
~(\ref{eqn:jac}) factor to yield
\begin{equation}
\label{eqn:fac1}
(1-\gamma_{\rho}^4)^3
(\gamma_{\rho}^6
-\gamma_{\rho}^2
+\frac{2}{3}\alpha^2)
=0
\end{equation}
\begin{equation}
\label{eqn:fac2}
{\rm and}\,\,\,\,\, 
J=
-(1-\gamma_{\rho}^4)^3
(\gamma_{\rho}^{10}
-\gamma_{\rho}^6
+\frac{1}{3}\alpha^2\gamma_{\rho}^4
+\frac{1}{9}\alpha^2)
\, ,
\end{equation}
respectively.  $\gamma_{\rho}=1$, which was already discarded in the
initial simplification leading to Eqs.~(\ref{eqn:root1}) and~(\ref{eqn:root2}), corresponds to the trivial linear
solution for the non-interacting case $\alpha=0$, since $\gamma_{\rho}\equiv
l_{\rho}/\sigma_{\rho}$.  The remaining factor in Eq.~(\ref{eqn:fac1}), a polynomial of
order three, may be solved
analytically to yield a local minimum of the $\epsilon_{{\rm GP}}$ surface:
\begin{eqnarray}
\label{eqn:length1}
&&\gamma_{\rho{\rm o}}
=\frac{1}{3^{1/4}}
\left\{2\cos\left[\frac{1}{3}\cos^{-1}
(-\sqrt{3}\,\zeta)\right]\right\}^{1/2}
\\
\label{eqn:length2}
&&\gamma_{z{\rm o}}
=\frac{2}{\sqrt{3\,\zeta}}
\cos\left[\frac{1}{3}\cos^{-1}(-\sqrt{3}\,\zeta)\right]\, ,
\end{eqnarray}
where $\zeta\equiv N^2 a^2/\sigma_{\rho}^2$ is the dimensionality
parameter, as explained below.  The shape of the $\epsilon_{{\rm GP}}$
surface is a bowl with a chute cascading down one side, a formation
often found in high mountains: the bowl represent a stability basin
and the chute, collapse.  A second non-trivial root of
Eq.~(\ref{eqn:fac1}) yields the saddle point at the start of the
chute, while a third root is extraneous.  The critical
point for collapse, which is also the $x$-intercept in Fig.~\ref{fig:crit}, is contained implicitly in the argument of the
arccosines in Eqs.~(\ref{eqn:length1}) and~(\ref{eqn:length2}),
\begin{equation}
\label{eqn:crit3}
-N a/\sigma_{\rho}\leq \frac{1}{3^{1/4}}\simeq 0.7598 \, .
\end{equation}
This expression differs by only a few percent from the
most extreme case for an expulsive potential, Eq.~(\ref{eqn:crit1}).
The value determined numerically by imaginary time relaxation of Eq.~(\ref{eqn:gpe3d}) is $-N
a/\sigma_{\rho}=0.6268 \pm 0.0035$, as illustrated in Fig.~\ref{fig:crit}.

At the critical point the soliton has size $(\gamma_{\rho{\rm
o}},\gamma_{z{\rm o}})=(3^{-1/4},3^{-1/4})$, from which it is clear
that it is three-dimensional.  Far from the collapse regime,
{\it i.e.}, for $\zeta\ll 1$, Eqs.~(\ref{eqn:length1})
and~(\ref{eqn:length2}) may be Taylor expanded to yield
\begin{eqnarray}
\label{eqn:expand1}
\gamma_{\rho{\rm o}} =
1-\frac{\zeta}{6}-\frac{7\,\zeta^2}{72}-\frac{13\,\zeta^3}{144}
+{\cal O}(\zeta^4)\, ,
\\
\label{eqn:expand2}
\gamma_{z{\rm o}} =
\frac{1}{\sqrt{\zeta}}
\left[1-\frac{\zeta}{3}-\frac{\zeta^2}{6}-\frac{4\,\zeta^3}{27}
+{\cal O}(\zeta^4)\right]\, .
\end{eqnarray}
For $\zeta\ll 1$, $\gamma_{z{\rm o}} \gg \gamma_{\rho{\rm
o}}$ and the soliton is quasi-one-dimensional.  $\zeta$ therefore
characterizes the dimensionality of the system.  

\subsection{Case III: The Quasi-One-Dimensional Limit}
\label{ssec:q1d}

For $\zeta\ll 1$ the axial harmonic degrees of freedom are not populated and the
system is essentially one-dimensional:  as shown in
Sec.~\ref{ssec:free}, $\gamma_{\rho}\sim 1$ and $\gamma_{z}\gg 1$.
For
strong transverse harmonic confinement it is
therefore a reasonable approximation to assume a wavefunction
of the form $\psi(\vec{r},t)=\phi(z,t)\exp[-(x^2+y^2)/(2\, \sigma_{\rho}^2)]$ and
integrate the GPE over the transverse degrees of freedom.  The
result is called the quasi-one-dimensional GPE, to first order in $\zeta$
\begin{eqnarray}
\label{eqn:gpe1d}
\left[-\frac{\hbar^2}{2m}\frac{\partial^2}{\partial z^2}
+g_{{\rm 1D}} N
\left|\phi(z,t)\right|^2
+ \hbar\omega_{\rho} + V(z)\right]\phi(z,t)
\nonumber \\
=i\hbar\,\frac{\partial}{\partial t}\,\phi(z,t)\, ,
\,\,\,\,\,
\end{eqnarray}
where $g_{{\rm 1D}}\equiv 2\,a\,\omega_{\rho}\hbar$ is the renormalized quasi-1D
coupling constant~\cite{castin2}.  As in Eq.~(\ref{eqn:gpe3d}), the
wavefunction has been normalized to one.  The constant factor of $\hbar\omega_{\rho}$ is the shift in energy due to the tranverse harmonic
confinement, $2\times\frac{1}{2}\hbar\omega_{\rho}$.

For $V(z)=0$, the analytically derived criterion used here for the
system to be in the quasi-one-dimensional regime is
equivalent to the intuitive criterion that the additional energy due
to atomic interactions is much less than the characteristic energy
scale of the transverse confinement, {\it i.e.},
$|\mu-\hbar\omega_{\rho}|\ll\hbar\omega_{\rho}$, where $\mu$ is the
chemical potential.  Consider the general
stationary solitonic
solution to Eq.~(\ref{eqn:gpe1d}), 
\begin{equation}
\label{eqn:sol}
\phi(z,t)=\frac{1}{\sqrt{2 l_z}}
{\rm sech}\left(\frac{z}{l_z}\right)\exp(- i\,\mu\,t/\hbar)\, .
\end{equation}
Assuming the transverse portion of the 3D wavefunction $\psi$ remains
in the linear ground state, Eq.~(\ref{eqn:sol}) may be substituted
into Eq.~(\ref{eqn:gpe1d}) to yield
$\mu=\hbar\omega_{\rho}-N^2g_{{\rm 1D}}^2 m/(8\hbar^2)=\hbar\omega_{\rho}-\zeta\hbar\omega_{\rho}/2$ and $l_z=2\hbar^2/(m |g_{{\rm 1D}}|
N)$.  Then $|\mu-\hbar\omega_{\rho}| \ll \hbar\omega_{\rho}$ implies
$|\zeta\hbar\omega_{\rho}/2|\ll\hbar\omega_{\rho}$, so that $\zeta/2\ll 1$.  The
intuitive geometrical criterion $l_{\rho}\ll l_z$ obtains essentially the same result: $l_{\rho}\ll
l_z = 2 \hbar^2/(m |g_{{\rm 1D}}| N)$, so that $\sqrt{\zeta}/2\ll 1$.
The same argument applies for $V(z)\neq 0$ provided that its effect on
the soliton length parameter is small.  In the
case where the characteristic transverse confinement length approaches the
$s$-wave scattering length $a$, the GPE, Eq.~(\ref{eqn:gpe1d}), no longer models the system and a
one-dimensional field theory with the appropriate effective coupling
constant must be considered instead~\cite{olshanii1}.
Since $|a|$ is on the order of angstroms for a stable attractive BEC,
this regime is not relevant to the present study.

Equation~(\ref{eqn:sol}) is a good ansatz to obtain the
stationary properties of a solitonic solution of
Eq.~(\ref{eqn:gpe1d}).  Consider the experimentally relevant case of
$V(z)=\frac{1}{2}m\omega_z^2z^2$, where $\zeta$ is retained to order
one. Then, using the
same variational methods as described in Sec.~\ref{ssec:3d}, one obtains
the following expression for the extrema:
\begin{equation}
\label{eqn:extrema1d}
\frac{\pi^2}{4}\lambda^2\gamma_z^4+\sqrt{\zeta}\gamma_z-1=0 \, .
\end{equation}
For $\omega_z^2>0$ there is a single real, positive root, which
represents a minimum of the energy functional; for $\omega_z^2<0$
there are two real, positive roots, which signify a minimum and a
maximum.  The minimum in the latter case,
the analytic form of which shall be needed in Sec.~\ref{ssec:evaporation}, is
written explicitly as $\gamma_z=F/\sqrt{\zeta}$, where $F\equiv
F(\zeta/|\lambda|)$ decreases monotonically from $\frac{4}{3}$ to 1
with increasing
$\zeta/|\lambda|$, and is defined as follows:
\begin{eqnarray}
\label{eqn:q1droot}
&F\equiv&\sqrt{\frac{G}{2}}
-\frac{1}{2}\sqrt{-2G+\frac{4\sqrt{2}}{\pi^2\sqrt{G}}\left(\frac{\zeta}{|\lambda|}\right)^2}\, ,
\\
&G\equiv &
\left\{\frac{4}{3\pi^{2/3}[1+\sqrt{1-(64\pi^2/27)(|\lambda|/\zeta)^2}]^{1/3}}\right\}
\left(\frac{\zeta}{|\lambda|}\right)^{2/3}
\nonumber \\
&&+\left\{\frac{[1+\sqrt{1-(64\pi^2/27)(|\lambda|/\zeta)^2}]^{1/3}}{\pi^{4/3}}\right\}
\left(\frac{\zeta}{|\lambda|}\right)^{4/3}
\, .
\nonumber
\end{eqnarray}

One obtains a
critical point for explosion, either from the Jacobean or directly
from the analytic form of the roots.  To prevent this from occurring, the parameters must be chosen
such that
\begin{equation}
\label{eqn:crit4}
\frac{|g_{{\rm 1D}}| N}{ \sqrt{m|\omega_z|/\hbar^3}}=2\,\sqrt{\frac{\zeta}{|\lambda|}}>
\left(\frac{2^{10}\,\pi^2}{3^3}\right)^{1/4}\simeq 4.3985\, .
\end{equation}
This quasi-1D prediction for stability against explosion is
illustrated in Fig.~\ref{fig:crit} (dashed line).  Since $l_z \simeq 2
\hbar^2/(m |g_{{\rm 1D}}| N)$, this criterion implies that the
longitudinal harmonic oscillator length for an expulsive potential
must be at least about two times the soliton width.  As shall be shown in
Sec.~\ref{ssec:evaporation}, quantum evaporation places a further
constraint on the available parameter space.

In the case where $\sqrt{|\lambda|/\zeta}$ is small, {\it i.e.}, far from
the critical value for explosion, Eq.~(\ref{eqn:q1droot}) may be Taylor expanded to obtain a simple analytic
expression for the soliton length:
\begin{equation}
\label{eqn:length1d}
\gamma_z=\sqrt{\frac{1}{\zeta}}\left[1
+\frac{\pi^2}{4}\frac{|\lambda|^2}{\zeta^2}
+\frac{\pi^4}{4}\frac{|\lambda|^4}{\zeta^4}
+{\cal O}\left(\frac{|\lambda|^6}{\zeta^6}\right)\right]
\, .
\end{equation}
To fourth order in $\sqrt{|\lambda|/\zeta}$, this expansion is accurate to within 1\% over half the permissible
parameter range given by Eq.~(\ref{eqn:crit4}).

\section{Nonlinear Dynamics}
\label{sec:nonlinear}

The dynamics of a bright soliton in an expulsive harmonic potential are
herein studied, to the authors' knowledge, for the first time.  Several intriguing new features
arise.  Firstly, sudden changes in trap parameters, such as was the case in
Ref.~\cite{carr29}, cause particle loss and shape oscillations.
Changing the strength of the nonlinearity via a Feshbach resonance has
similar effects.  In both cases the excitations are demonstrated to damp exponentially for
small changes.  Secondly, it
is shown that a large change in the nonlinearity can be used, with present experimental apparatus, to create a higher order
soliton.  Thirdly, the lifetime of a soliton in an expulsive potential
is not infinite.  The effective potential experienced by
an atom in the soliton is the external potential plus the
potential well created by the mean field, itself a function of the
number of atoms.  Thus the wavefunction undergoes tunneling with a
rate which depends on $N$.
This nonlinear process accelerates, eventually leading to
explosion.  

The exponential damping of weak excitations, nonlinear tunneling, and
eventual explosion of a soliton differ radically from the case of a
constant potential.  In the following sections these phenomena are studied
analytically and numerically.  In the latter case the condensate wavefunction, which solves Eqs.~(\ref{eqn:gpe3d})
and~(\ref{eqn:gpe1d}) in three and one dimensions, respectively, is
expanded as $\psi(\vec{r},t)=\Sigma_j\,c_j(t)\Phi_j(\vec{r})$.  The
$\Phi_j(\vec{r})$ are used as a coordinate-discretized pseudo-spectral
basis, and the $c_j(t)$ are propagated in time with a fourth-order
variable step Runge-Kutta algorithm.  These numerical techniques are
standard~\cite{press1}.  Absorbing bounds,
{\it i.e.}, imaginary potential wells, are added far from the soliton
in order to simulate an unbounded system, with care being taken to insure there
are no reflections~\cite{note1}.

The motion of the center of mass of the condensate is decoupled
from the relative motion of its constituent atoms, as is generically
the case for harmonic potentials.  The center of mass is unstable in
an expulsive potential; in the following only the relative
motion of the atoms is considered.  Thus the condensate wavefunction is symmetrized in our
simulations in order to keep its center of mass at rest.

\subsection{Non-equilibrium Soliton Evolution in Three Dimensions}
\label{ssec:motion}

In order to
understand the time evolution of a soliton which is created in a
non-equilibrium initial state, one may begin with the equations of motion
for the variational parameters of Eq.~(\ref{eqn:ansatz3d}).  In
general, one adds a quadratic imaginary factor
in $x$, $y$, and $z$ to the exponential in Eq.~(\ref{eqn:ansatz3d})
and then performs a Lagrangian variational
analysis~\cite{desaix1,perez2}.  Using this method, one finds
\begin{equation}
\label{eqn:ham1}
\ddot{\gamma}_{\rho}
=\frac{1}{\gamma_{\rho}^3}-\gamma_{\rho}
+\frac{2\alpha}{3\gamma_{\rho}^3\gamma_z} \, ,
\end{equation}
\begin{equation}
\label{eqn:ham2}
\left(\frac{\pi^2}{12}\right)\ddot{\gamma}_{z}
=\frac{1}{3\gamma_{z}^3}-\frac{\pi^2}{12}\gamma_{z}\lambda^2
+\frac{\alpha}{3\gamma_{z}^2\gamma_{\rho}^2} \, ,
\end{equation}
where the derivative has been taken with respect to rescaled time
$\tau\equiv\omega_{\rho} t$.  These equations match those of
Ref.~\cite{perez2}, up to small numerical factors due to the asymmetry
between the hyperbolic secant and gaussian functions in the ansatz.
The factor of $\pi^2/12$ in Eq.~(\ref{eqn:ham2}) has been made
explicit in order to demonstrate the difference between the Lagrangian
analysis and simply taking
the first derivative of Eq.~(\ref{eqn:energy2}), which gives an
identical result up to a slight difference in the effective mass.
One may then
perform a linear perturbation analysis around the minimum and derive
small excitation frequencies.  Since this has already been done
elsewhere~\cite{singh1,perez2} it is not repeated here.  

%%%%%%%%%%% figure 2 %%%%%%%%%%%
%
\begin{figure}[hb]
\begin{center}
\epsfxsize=8cm \leavevmode \epsfbox{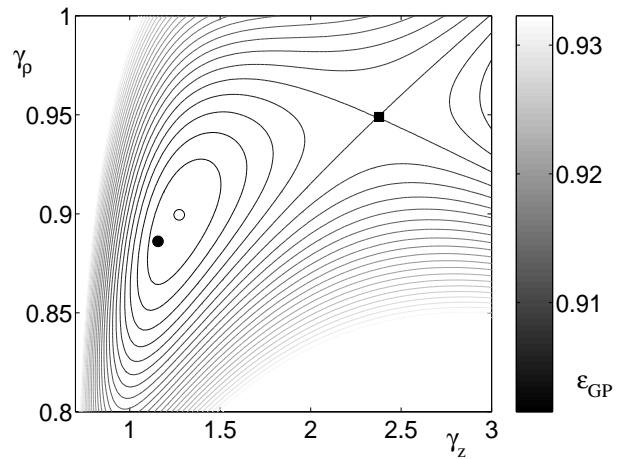}
\caption{\label{fig:contour}  Contour plot of energy surface
for a matter-wave bright soliton in an expulsive potential.  The
parameters of Ref.~\cite{carr29} have been used, with $Na/\sigma_{\rho}=-0.6750$, $\omega_z = 2\pi i \times 70$ Hz and
$\omega_{\rho}=2\pi \times 700$ Hz.  The square
marks the lower of the saddle points and the closed contour to the
left within which a
stable soliton may be formed; the open circle marks the local minimum;
and the filled circle marks the initial conditions of
Ref.~\cite{carr29}: the soliton was projected from an
attractive harmonic potential $\omega_{zi}= 2\pi \times 100$ Hz.}
\end{center}
\end{figure}
%
%%%%%%%%%%%%%%%%%%%%%%%%%%%%%%%%%%%%%%%%

The essential limitations on changes in the
soliton and trapping parameters to avoid collapse and/or explosion can be gleaned directly from the
variational studies of Sec.~\ref{sec:stability}.  The shape
of the energy surface $\epsilon_{{\rm GP}}(\gamma_{\rho},\gamma_z)$ is
that of a long, narrow rift running along $\gamma_z$.  The sides of the rift slope rapidly
upwards towards infinity.  Towards increasing $\gamma_z$ there is a
saddle point beyond which explosion occurs; towards decreasing
$\gamma_z$ the rift turns to point toward the origin and ends at a
second saddle point, after which the condensate collapses.  The turning is due to
the three-dimensional nature of the soliton near collapse.  Let the lower of these two saddle
points have an energy $\epsilon_{{\rm max}}$.  Then the curve
$\epsilon_{{\rm GP}}(\gamma_{\rho},\gamma_z)=\epsilon_{{\rm max}}$
defines a closed loop which signifies the lower bound for creation of
a soliton.  Any soliton created with the initial condition
$\dot{\gamma}_{\rho}=0$ and $\dot{\gamma}_{z}=0$ which has length and width within this loop, though its motion along the
energy surface may be complicated, must, by
conservation of energy, be stable against collapse and explosion.  In
Fig.~\ref{fig:contour} is shown an example of the energy surface for
the case of Ref.~\cite{carr29}.  The loop defined by $\epsilon_{{\rm
max}}$ is marked with a square, while the open and filled circles
represent the minimum and the initial state, respectively.

The above variational arguments do
not include phonon or particle emission.  Thus they cannot  model the damping
of excitations, which, as shall be illustrated in the following section is,
for an expulsive harmonic potential, exponential.

\subsection{Damping of Soliton Excitations in One Dimension}
\label{ssec:project}

Bright solitons in fiber optics are well known to shed their
excess power by radiation.  This is an example of the robustness which
makes them so useful for long distance communications.
Thus any pulse injected with an area of
$\pi/2<\int_{-\infty}^{+\infty}dz\Upsilon(z,0)<3\pi/2$ into a fiber optic
results in a soliton with amplitude
oscillations which decay as $1/\sqrt{t}$, where the governing
one-dimensional focusing nonlinear Schr\"odinger equation 
is conventionally written as $i\partial_t \Upsilon =
-\frac{1}{2}\partial_z^2\Upsilon +|\Upsilon|^2\Upsilon$ without a fixed normalization~\cite{haus1}.  A sudden change in the scattering length
or number of particles in  a matter-wave bright soliton is, by a
rescaling of the wavefunction amplitude, time, and position,
mathematically identical to
injecting a soliton of the form $\Upsilon(z,0)=A\,{\rm sech}(z)$ into the
above equation, where $A\equiv \sqrt{\eta_f/\eta_i}$ with
\begin{equation}
\label{eqn:eta}
\eta\equiv 2N|a|/\sigma_{\rho}^2=|g_{{\rm 1D}}| N m/\hbar^2
\end{equation}
from
Eq.~(\ref{eqn:gpe1d}).  The subscripts $i$ and $f$ signify the states
just before and just after the abrupt change in the scattering length.  This case has been solved previously with
the inverse scattering transform~\cite{satsuma1}.  Supposing $A=1+b$
with $|b|<\frac{1}{2}$, the resulting number of particles $N_p$ emitted from
a soliton with initial number of particles $N_{i}$ is 
\begin{equation}
\label{eqn:num1}
N_p =N_{i}\left(1-\sqrt{\frac{\eta_f}{\eta_i}}\,\right)^2\frac{\eta_i}{\eta_f}
\end{equation}
Supposing $\eta_f/\eta_i=1+\delta$, where $|\delta|\ll 1$, one finds $N_p=N_{i}\times \delta^2/4$, from which it is apparent that losses are very
small indeed.  

An identical
result can be obtained by the overlap
integral between the initial and final soliton
wavefunctions~\cite{castin3}, as is now demonstrated.  Let $\phi_i(z)$
be the initial wavefunction of the soliton with parameter $\eta_i$,
and let $\phi_f (z)$ be the wavefunction of the steady state soliton 
with parameter $\eta_f$.  One can always split the initial wavefunction
$\phi_i$ into a component parallel to $\phi_f$ and a component 
$\delta\phi_{\perp}$ orthogonal to $\phi_f$, so that
$\phi_i(z) = \sqrt{1-\epsilon}\,\phi_f(z) + \delta\phi_{\perp}(z,0)$.
Since $\epsilon \ll 1$ for $\eta_f$ close to $\eta_i$,
$\delta\phi_{\perp}$ has a small norm $\sqrt{\epsilon}$ and one may
calculate its temporal evolution by linearizing the GPE around the steady state solution
$\phi_f$.  The spectrum of linear excitations, also called the
Bogoliubov spectrum, is $E_k=\hbar^2k^2/(2m)+|\mu_f|$, where $k$ is the
wavevector of the emitted wave and $\mu_f$ is the chemical potential
of the depleted soliton at long times~\cite{kaup1,olshaniicommunication1}.  As a consequence of this dispersion relation, the emitted wave $\delta\phi_{\perp}$ will spread as an unconfined wavepacket, with
a spatial width which scales as $t$ and a maximum amplitude 
as $1/\sqrt{t}$. Therefore the number of particles emitted by the soliton
at long times is simply the number of particles in the component 
$\delta\phi_{\perp}$, that is, $N_p=N_i(1-|\langle \phi_f|\phi_i\rangle|^2)$.  The
overlap integral is
\begin{equation}
\label{eqn:overlap}
\langle \phi_f|\phi_i\rangle=
\langle \phi(z;\eta_f)|\phi(z;\eta_i)\rangle
\simeq 1-\delta^2/8\, 
\end{equation}
to second order in $\delta$.  It follows directly that $N_p\simeq
N_i\times \delta^2/4$, as was found above by expanding
Eq.~(\ref{eqn:num1}) in $\delta$.

In Ref.~\cite{carr29} the abrupt change in parameters in
Eq.~(\ref{eqn:gpe3d}) occurs in the longitudinal trapping frequency
$\omega_z$, rather than the strength of the mean field.  Although this
case is not covered by the inverse scattering
transform above, one may generalize the considerations given by the
overlap integral of Eq.~(\ref{eqn:overlap}).  One must recalculate the
linear excitation spectrum in the presence of an expulsive harmonic
potential.  To obtain a discrete spectrum the system is treated
as enclosed in a box of length much larger than the soliton size;
periodic boundary conditions are chosen.  Neglecting quantum
reflections, the eigenmodes may be treated as semi-classical plane waves.  The effect
of the soliton on these waves is simply a shift in phase, similiar to
the case of a free soliton~\cite{kaup1}, as may be
seen by the following.  

Consider a
semi-classical wave with wavevector $k$ emerging from the soliton.
The expulsive potential can be neglected over the width
of the soliton provided that $|\mu|\gg \hbar|\omega_z|$, {\i.e.}, 
$|\lambda|/\zeta \ll 1/5$, so that the soliton is very far from explosion,
as given by Eq.~(\ref{eqn:crit4}).  Near the edge of the soliton one
must verify the applicability of the WKB approximation with the
standard criterion $|d/dz(k(z)^{-1})|\ll 1$.  Given
$k=\sqrt{E+|\omega_z|^2z^2/2}$, $|d/dz(k(z)^{-1})|= (|\omega_z|^2
z/2)/(E+|\omega_z|^2z^2/2)^{3/2}$.  Since the soliton edge is being
considered,  $z\simeq l_z$, and since the emerging wave is created by
a global change in the soliton shape, $k\simeq 1/l_z$ so that $E\simeq \hbar^2/(2ml_z^2)$.  This leads to the same condition as above, $|\mu|\gg \hbar|\omega_z|$.  In this case the
transmission coefficient is $T=(2k/\eta+i)^2/(2k/\eta-i)^2=\exp(i\theta)$ , where $\theta$ varies from $2\pi$ to zero as $2k/\eta$ varies from zero to
infinity~\cite{kaup1,olshaniicommunication1}.  This shift in phase does not change the density of states
in the limit where the box length tends to infinity.  One concludes that the expansion of linear excitations is unaffected by
the soliton.  Thus the spatial width of $\delta\phi_{\perp}$ increases
as $\exp(|\omega_z|t)$ and its amplitude decays as
$\exp(-|\omega_z|t/2)$ for times long as compared to $1/|\omega_z|$.  Equation~(\ref{eqn:overlap}) gives
\begin{equation}
\label{eqn:loss}
N_p \simeq \frac{\pi^4}{2^6\zeta^4}\left(|\lambda_i|^2-|\lambda_f|^2\right)^2
\end{equation}

\begin{figure}[hb]
\begin{center}
\epsfxsize=8cm \leavevmode \epsfbox{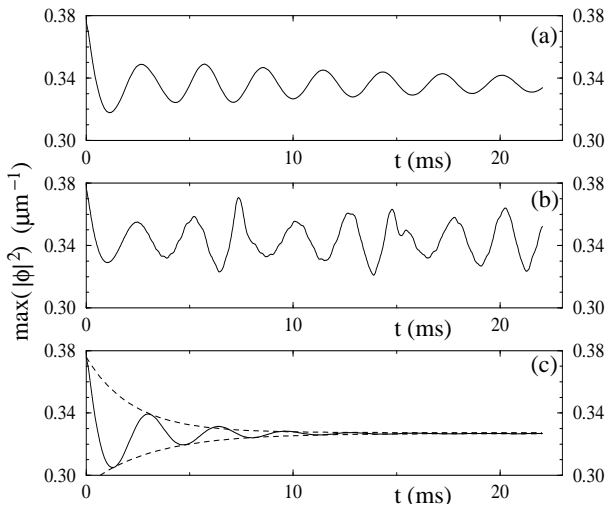}
\caption{\label{fig:shed}  Fluctuations in the peak value of the
quasi-1D wavefunction density for a 10\% change in nonlinearity from $\eta_i=-1.5\,\mu{\rm m}^{-1}$ to $\eta_f=-1.35\,\mu{\rm m}^{-1}$ are shown for (a) a
constant potential, (b) an
attractive harmonic potential of frequency $\omega_z= 2\pi \times 65$ Hz, and (c) an
expulsive harmonic potential of frequency $\omega_z=
2\pi i \times 65$ Hz (solid line).  (a) damps as $1/\sqrt{t}$, similar to a
wavepacket, (b) recurs, and (c)
damps exponentially as $\exp(-\kappa t)$, where
$\kappa\simeq|\omega_z|$: $\kappa=72\times 2\pi$ Hz and $51\times 2\pi$ Hz for the upper and
lower exponential fits to the decaying envelope, respectively (dashed
lines).  The mass of $^7$Li has been used to obtain the time scale.}
\end{center}
\end{figure}

In Fig.~\ref{fig:shed}
the cases of free and expulsive harmonic potentials
discussed above as well as that of an attractive harmonic potential
are illustrated.
Numerical integration of the quasi-1D GPE shows that a 10\% change in the value of $\eta$
results in irregular oscillations for the attractive potential, damping of the
maximum density as $1/\sqrt{t}$ for zero external potential,
and damping as $\exp(-\kappa t)$ for the expulsive potential, where
$\kappa\simeq|\omega_z|$.  In the latter case, Fig.~\ref{fig:shed}(c), $N_p\simeq N_{i}\times 0.0031$, whereas the prediction from above is $\sim
0.0014$.  This enhanced loss is due to nonlinear tunneling,
as shall be explained in Sec.~\ref{ssec:evaporation}.  In simulations $N_p$ 
is determined by evaluating the norm squared of the wavefunction as a
function of time, since the absorbing bounds remove atoms
outside the immediate region of the soliton.   

%%%%%%%%%%% figure 3 %%%%%%%%%%%
%
%
%%%%%%%%%%%%%%%%%%%%%%%%%%%%%%%%%%%%%%%%

Thus, provided that the soliton is created with width and length dimensions that lie within the bounds obtained from variational
studies of the final parameter set,  one may expect it to shed
in a time $1/|\omega_z|$ its excess mass and come to rest at the local minimum of the energy
surface.

\subsection{Higher Order Soliton Creation}
\label{ssec:higher}

In the preceeding section, the matter-wave parallel to the injection of
an optical soliton of the form $\Upsilon(z,t=0)=A\,{\rm sech}(z)$ with
$A=1+b$ and $|b|<\frac{1}{2}$ was considered.  A soliton of this form
is in fact the first of a denumerably infinite family of higher order
soliton solutions to the one-dimensional focusing nonlinear Schr\"odinger
equation.  Each such solution can
be produced from the initial state $A=n$, where $n$ is a positive
integer, without loss of particles.  Moreover, provided that
$|b|<\frac{1}{2}$, $A=n+b$ will always result in a soliton of order
$n$, with a fractional loss of $N_p/N_i=b^2/(n+b)^2$, as derived in Ref.~\cite{satsuma1}.  Bright solitons with zero relative phase have an attractive
interaction.  As a soliton of the form $\Upsilon(z,t=0)=n\,{\rm
sech}(z)$ has less than $n$ times the
energy of a soliton of form $\Upsilon(z,t=0)=\,{\rm sech}(z)$, a
higher order soliton can be interpreted as tightly bound state of $n$
strongly overlapping solitons.  It is periodic in time in both its phase and density.

A {\it matter-wave} higher order soliton can be created by a sudden shift in the
scattering length.  In particular, the order two soliton requires that
the
scattering length be decreased by a factor of four: for $^7$Li this factor can be produced
simply for the state $|F=1,m_F=1\rangle$ by a shift in magnetic field
from $\sim 175$ gauss to $\sim 275$ gauss, thereby shifting $a$ from
$-a_0$ to $-4a_0$, where $a_0$ is the Bohr radius.  As with the soliton of order
one, higher order solitons persist in the presence of a sufficiently weak expulsive
potential.

One supposes two experimental approaches: in the first, a
soliton, already formed in an expulsive potential, is subjected to a
sudden change in the scattering length; alternatively, the scattering
length of an attractive
condensate trapped in an attractive harmonic potential is
simultaneously shifted by a factor of four with the projection of the soliton onto a
weakly expulsive potential.  The latter may be easier for observing
oscillations of the soliton shape, since the center of mass in
unstable unless the relative position of the potential and the soliton
are continuously adjusted.  In either case the initial scattering
length must be chosen such that it satisfies Eq.~(\ref{eqn:crit4}),
lest the soliton explode as it returns periodically to the form of its
initial state.  Thus, to avoid both collapse and explosion on experimental time scales
and simultaneously accomodate the factor
of four change in $\eta$, one should choose trapping
potentials such that $|\omega_z/\omega_{\rho}|^{1/2} < 0.1$, as deduced from Fig.~\ref{fig:crit}.

\begin{figure}[hb]
\begin{center}
\epsfxsize=8cm \leavevmode \epsfbox{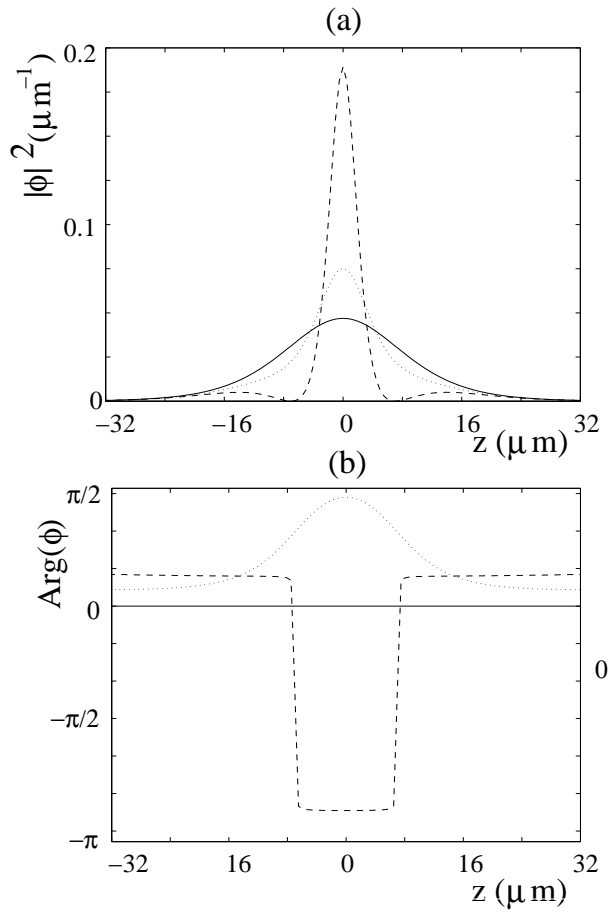}
\caption{\label{fig:hos}  The 1D numerical evolution of the (a) density and (b) phase of a higher order
soliton created by an abrupt change in scattering length by a factor
of four.  The
oscillation period is 19 ms for $^7$Li, with the initial state (solid
line), quarter period (dotted line), and half period (dashed line)
shown.  An
expulsive potential hill of frequency $\omega_z=2\pi i\times 5$ Hz is
present in the simulation, with $\eta_i=0.187 \,\mu{\rm m}^{-1}$ and $\eta_f=0.750 \,\mu{\rm m}^{-1}$.
Supposing a transverse oscillator frequency of
$\omega_{\rho}=2\pi\times1000$ Hz and
a change in scattering length from $a_i=-a_0$ to $a_f=-4 a_0$, the
resulting soliton consists of 2550 atoms and is well within the quasi-one-dimensional regime.  Note the phase jump of $\pi$ in the phase corresponding with the spontaneous
development of nodes in the density.  Such a nonlinear time-dependent structure could be created and observed in
present BEC experiments.}
\end{center}
\end{figure}

The order two soliton solution to Eq.~(\ref{eqn:gpe1d}) with $V(z)=0$ takes the form~\cite{satsuma1}
\begin{eqnarray}
\label{eqn:2sol}
&&\phi(z,t)=\frac{\sqrt{\eta_f}}{2}\,
\exp\left(i \frac{\hbar\eta_f^{2} t}{128 m}\right)
\nonumber \\
&&\times\left[\frac{
3\exp\left(i \frac{\hbar\eta_f^{2} t}{16 m}\right)
+{\rm cosh}\left(\frac{\eta_f z}{8}\right)
+{\rm cosh}\left(\frac{3 \eta_f z}{8}\right)
}
{
3\cos\left(\frac{\hbar\eta_f^{2} t}{16 m}\right)
+4\,{\rm cosh}\left(\frac{\eta_f z}{4}\right)
+{\rm cosh}\left(\frac{\eta_f z}{2}\right)
}\right]
\end{eqnarray}
where $\eta_f\equiv 2 N |a_f| /\sigma_{\rho}^2$, as above, with $a_f$ being the final
scattering length.  Thus the period is $16 m/(2\pi \hbar \eta_f^{2})$
while the spatial width is on the order of $8/\eta_f$.
Figure~\ref{fig:hos} illustrates the creation of an order two soliton
by abruptly changing the scattering length from $\eta_i=-0.1875\,\mu{\rm m}^{-1}$ to
$\eta_f=-0.75\,\mu{\rm m}^{-1}$ and
simultaneously projecting the soliton from a constant to an expulsive
potential of frequency $\omega_z= 2\pi\,i\times 5$ Hz.  This matches the analytic form of
Eq.~(\ref{eqn:2sol}).  A negligible number of particles $N_p/N_{i}\simeq 0.001\%$ are emitted due to the projection, in agreement with
the prediction of Eq.~(\ref{eqn:loss}), here extended to the case $n=2$.  The above simulated parameters
are equivalent to an experiment for which $\omega_{\rho}=1000\times 2\pi$
Hz, $\omega_z=2\pi i\times 5 $ Hz, $a_i=-a_0$, $a_f=-4 a_0$, and $N=2550$
particles for $^7$Li.  The shape variations of this time-dependent
nonlinear structure should be observable {\it in situ} with a period
of 19 ms.

%%%%%%%%%%% figure 4 %%%%%%%%%%%
%
%
%%%%%%%%%%%%%%%%%%%%%%%%%%%%%%%%%%%%%%%%

\subsection{Quantum Evaporation}
\label{ssec:evaporation}

\begin{figure}[hb]
\begin{center}
\epsfxsize=8cm \leavevmode \epsfbox{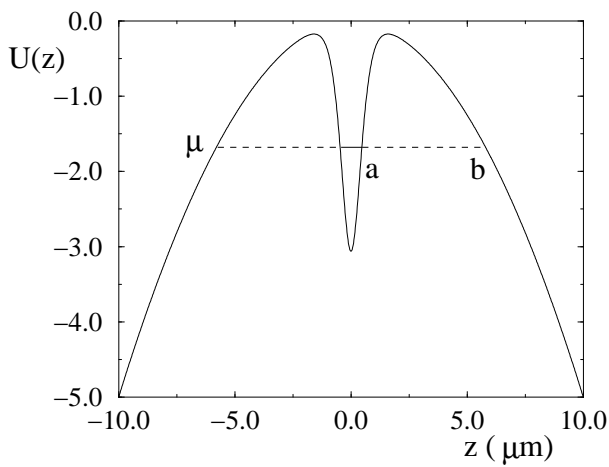}
\caption{\label{fig:pot}  The effective potential well experienced by a
bright soliton in a one-dimensional expulsive harmonic potential is
illustrated.  The energy of the single bound state is $\mu$, and the
wavefunction tunnels in either direction through a barrier of length $b-a$, as
indicated by the dashed lines. Here we have used the parameters
$\eta=-3.5\,\mu{\rm m}^{-1}$ with a steep potential hill of
$\omega_z=2\pi i\times 450$ Hz for illustration purposes.}
\end{center}
\end{figure}

A bright soliton in an expulsive potential experiences an effective
potential $U(z,t)=m\omega_z^2 z^2/2+g_{{\rm 1D}}N|\phi(z,t)|^2$, as
sketched in Fig.~\ref{fig:pot}.  This is a time-dependent
potential well, the depth of which is a function of the number of particles.  The soliton therefore undergoes
nonlinear tunneling, which we have called quantum evaporation.  To
estimate the rate of quantum evaporation one may consider linear
tunneling in the external potential $U(z,t)$, where one neglects the
explicit time dependence of the wavefunction.  The
semi-classical WKB approximation then predicts, for a one-dimensional
system~\cite{landau2}, 
\begin{equation}
\label{eqn:wkb}
\Gamma=2\,\frac{\omega}{2\pi}\exp\left(-2\,\left|\int_a^b \frac{dz\,p(z)}{\hbar}\right|\right)\, ,
\end{equation}
where the first factor of 2 comes from the fact that the
wavefunction can tunnel in two directions, $\omega$ is a
characteristic oscillation frequency of the
particles in the well, $b-a$ is the barrier length through which the wavefunction must
tunnel, and $p(z)\equiv\sqrt{2m}\sqrt{\mu-U(z)}$, where $\mu$ is the
chemical potential.  Since the effective
potential well contains a single bound state (see Fig.~\ref{fig:pot}),
one may take heuristically
$\frac{1}{2}\hbar\omega=|\eta|\phi(0)|^2-\mu|$, which is the height of
the chemical potential above the minimum of the well.  

%%%%%%%%%%% figure 5 %%%%%%%%%%%
%
%
%%%%%%%%%%%%%%%%%%%%%%%%%%%%%%%%%%%%%%%%

Supposing that the
wavefunction is modeled by the variational ansatz Eq.~(\ref{eqn:sol}),
one may determine $\mu$ and the well depth $\eta|\phi(0)|^2$ directly
from the methods outlined in Sec.~\ref{ssec:q1d}: $\phi(z)$ is
given by Eq.~(\ref{eqn:sol}), the length $l_z=2F/\eta$ by the real
positive root of Eq.~(\ref{eqn:extrema1d}) for $\omega_z^2<0$, 
and the chemical potential
by subsequently minimizing the eigenvalue of Eq.~(\ref{eqn:gpe1d}).
The semi-classical WKB tunneling rate, scaled to the frequency of the expulsive potential, is
then given by
\begin{equation}
\label{eqn:wkb2}
\frac{\Gamma}{|\omega_z|}
=\frac{2}{\pi}\left(\frac{\zeta}{F|\lambda|}-\frac{|\mu|}{\hbar|\omega_z|}\right)\exp(-2I)\, ,
\end{equation}
where
\begin{equation}
\label{eqn:int}
I\equiv \int_{\tilde{a}}^{\tilde{b}} dx
\left|\frac{2F^2|\lambda|\mu}{\zeta\hbar|\omega_z|}+\frac{F^4|\lambda|^2}{\zeta^2}x^{2}+2F\,{\rm
sech}^2(x)\right|^{1/2}
\end{equation}
and
\begin{equation}
\label{eqn:cp}
\frac{\mu}{\hbar|\omega_z|}=\left(\frac{\zeta}{6|\lambda|F^2}
-\frac{2\,\zeta}{3|\lambda|F}
-\frac{\pi^2|\lambda|F^2}{24\,\zeta}\right)\, .
\end{equation}
Here the integration variable $x\equiv z/l_z$ and the parameters $\tilde{a}$
and $\tilde{b}$ are determined by $U(\tilde{a}),\,U(\tilde{b})=\mu$,
as illustrated in Fig.~\ref{fig:pot}.
The function $F= F(\zeta/|\lambda|)$, as defined by Eq.~(\ref{eqn:q1droot}), varies monotonically between 1 and
4/3, the latter value occuring at the critical point for explosion.
Here the additional
chemical potential $\hbar\omega_{\rho}$
due to transverse confinement is neglected, since in the quasi-1D
regime this is assumed to
have no effect on the dynamics.
Equation~(\ref{eqn:wkb2}) is a function of the
single parameter $\zeta/|\lambda|=N^2a^2\omega_{\rho}^2m/(\hbar|\omega_z|)$.

In the case where the soliton is far from explosion, {\it i.e.}, $\zeta/|\lambda|\gg\sqrt{2^6\pi^2/
3^{3}}\equiv(\zeta/|\lambda|)_c$, as given by Eq.~(\ref{eqn:crit4}),
the above reduces to the following simple limit:  $F\rightarrow 1^+$,
$\mu/(\hbar|\omega_z|)\rightarrow -\zeta/(2|\lambda|)$, and
$I\rightarrow\pi\zeta/(4|\lambda|)-\sqrt{2}\,{\rm
ln}(1+\sqrt{2})+{\cal O}(|\lambda|/\zeta)$.  The asymptotic tunneling rate is therefore
\begin{equation}
\label{eqn:wkb3}
\frac{\Gamma}{|\omega_z|}
\simeq\frac{2 (1+\sqrt{2})^{2\sqrt{2}}}{\pi}\frac{|\mu|}{\hbar|\omega_z|}
\exp\left(-\pi\frac{|\mu|}{\hbar|\omega_z|}\right)\, ,
\end{equation}
a function of the single parameter $\mu/(\hbar|\omega_z|)$.

%%%%%%%%%%% figure 6 %%%%%%%%%%%
%
\begin{figure}[hb]
\begin{center}
\epsfxsize=8cm \leavevmode \epsfbox{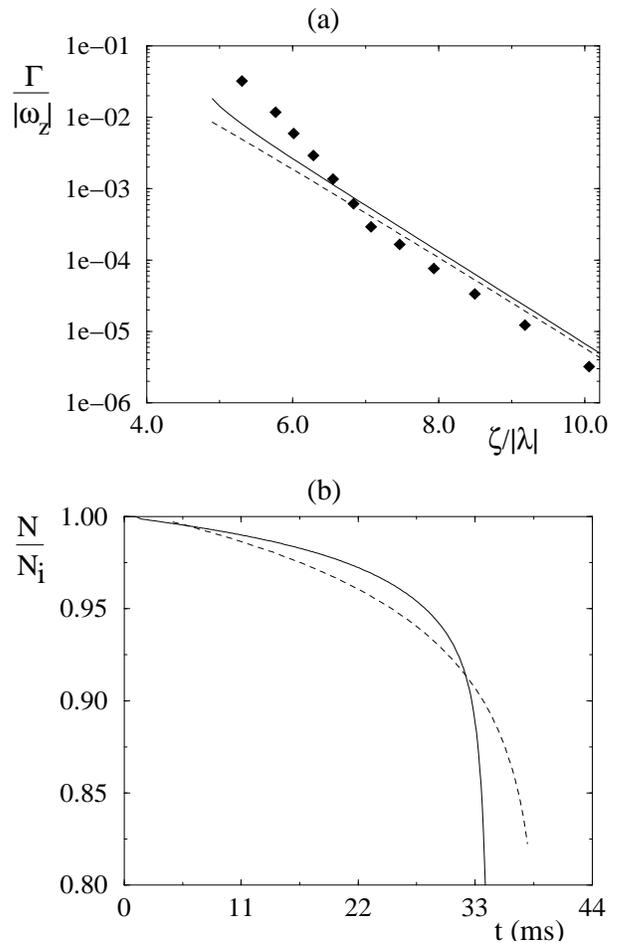}
\caption{\label{fig:qe2}  Quantum evaporation.  (a) The WKB prediction
for the tunneling rate as a function of
$\zeta/|\lambda|=N^2a^2|\omega_z|m/\hbar$ (solid line), and its large
$|\mu|/(\hbar|\omega_z|)$ asymptotic expression (dashed line).  Black diamonds: numerical
data produced by integration of the quasi-1D GPE.  (b) Solid line:
explosion of a matter-wave bright soliton in
the quasi-one-dimensional regime near
the critical point, obtained by numerical integration.  Dashed line: WKB prediction of the same.  Here
$\eta_i=-3.0 \, \mu{\rm m}^{-1}$ and $\omega_z = 2\pi i
\times 450$ Hz.  The mass of $^7$Li has been used to obtain the time
scale.}
\end{center}
\end{figure}
%
%%%%%%%%%%%%%%%%%%%%%%%%%%%%%%%%%%%%%%%%

In Fig.~\ref{fig:qe2}(a) is shown the complete and asymptotic
WKB
variational predictions of
Eq.~(\ref{eqn:wkb2}) (solid line) and Eq.~(\ref{eqn:wkb3}) (dashed
line), respectively, and numerical data obtained by direct
simulation of Eq.~(\ref{eqn:gpe1d}) (black diamonds).  In Fig.~\ref{fig:qe2}(b) the nonlinear tunneling of a soliton in an
expulsive harmonic potential is shown explicitly.  The solid line shows
the result given by numerical evolution of the quasi-1D GPE.  In this
simulation $(\zeta/|\lambda|)_i=7.1151$, or
$2\sqrt{(\zeta/|\lambda|)_i}=5.3348$.  It is apparent that the
tunneling accelerates and the soliton explodes, the latter occurring after about 5\% of the
particles in the soliton have been lost.  This differs by about 10\%
from the predicted critical point of Eq.~(\ref{eqn:crit4}), pointing to a level
of accuracy in the quasi-1D variational approximation similiar to that
of the 3D variational approximation of Sec.~\ref{ssec:3d} and
Fig.~\ref{fig:crit}.  The dashed line represents the numerical
integration of $dN/dt=-\Gamma\, N$, where $\Gamma$ is obtained by the
WKB approximation as shown in Fig.~\ref{fig:qe2}(a).  The explosion
time is similar to that of the simulation; the critical particle number is
as given by the variational approximation, as assumed in the
above derivation.

To
experimentally observe
quantum evaporation and subsequent explosion one faces the difficulty
that the center of mass of the soliton is unstable.  There are two solutions: one may
select parameters such that the tunneling rate is maximized and try to
observe the soliton as it falls; or one may
use a soliton which is large enough to observe {\it in situ} and
construct a feedback loop to continuously shift the peak of the expulsive potential
so as to keep the soliton centralized at its maximum.

Note that a soliton of order $n>1$ is expected to show the same kind of
tunneling properties.

\section{Conclusions}
\label{sec:conclusions}

The size, energy, and stability regimes of a stable matter-wave bright soliton
in a harmonic potential which is transversely attractive and
longitudinally expulsive have been presented.  The critical branches
for collapse and explosion have been delineated analytically.  The equations of
motion of the soliton parameters were calculated, and the response of the
soliton to sudden
changes in trap parameters or the scattering length was shown to
result in excitations which damp exponentially by emission of
particles from the soliton.  It was shown that a higher order
soliton can be created by a sudden change in the scattering length by
a factor of four.  Finally, an intriguing nonlinear dynamical effect was presented: a matter-wave bright soliton undergoes quantum evaporation
and eventually explodes.

The parameter space for creation of a locally stable matter-wave bright soliton is accessible to present experiments, as already demonstrated
in Refs.~\cite{carr29} and~\cite{strecker1}.  Data from the former shows that solitons are
self-adjusting: when the scattering length of a condensate with $\sim$20,000
atoms was tuned to be negative, a large thermal cloud with a stable condensate core of $\sim$5,000 atoms was observed to remain.  Projection onto an expulsive potential
caused the observed thermal cloud to be rapidly expelled from the vicinity of
the condensate, resulting in a nearly pure soliton.  The
choice of a weaker expulsive potential and a less negative scattering
length than that of Ref.~\cite{carr29} could be used to create
solitons observable {\it in situ}.  Thus the experimental outlook
for the investigation of matter-wave bright solitons is very positive.

An important aspect of solitons which could be investigated in future
work is their interactions: this is an essential aspect of solitons
which gives them their particle-like nature, in contrast to a solitary
wave.  Binary interactions between solitons depend on their relative
phase: for $\Delta\phi=0$ they are attractive, while for
$\Delta\phi=\pi$ they are repulsive~\cite{gordon1,carr19}.  Such interactions
could therefore measure phase decoherence times.

{\bf Acknowledgments:} We thank Iacopo Carusotto, 
Gora Shlyapnikov and the group of Christophe Salomon 
(Thomas Bourdel, Julien Cubizolles, Gabriele Ferrari, Lev Khaykovich, 
and Florian Schreck) for
helpful discussions.  This work was supported by the Distinguished
International Fellowship Program of the National Science Foundation
grant no. MPS-DRF 0104447. Laboratoire Kastler Brossel is
an unit\'e de recherche de l'\'Ecole normale sup\'erieure et de
l'Universit\'e
Pierre et Marie Curie, associ\'ee au CNRS.
%\bibliographystyle{acs}
%\bibliography{/users/lkb/lcarr/writing/refs/refs}

\begin{thebibliography}{10}

\bibitem{campbell1}
D. Campbell,  in {\em Complex Systems, SFI Studies in the Sciences of
  Complexity}, edited by D. Stein (Addison-Wesley Longman Publishing Group
  Ltd., New York, 1989), pp.\ 3--105.

\bibitem{hasegawa1}
A. Hasegawa, {\em Optical Solitons in Fibers} (Springer-Verlag, New York,
  1990).

\bibitem{agrawal1}
G.~P. Agrawal, {\em Nonlinear Fiber Optics}, 2nd ed. (Academic Press, San
  Diego, 1995).

\bibitem{denschlag1}
J.~Denschlag, J.~E. Simsarian, D.~L. Feder, C.~W. Clark, L.~A. Collins,
  J.~Cubizolles, L.~Deng, E.~W. Hagley, K.~Helmerson, W.~P. Reinhardt, S.~L.
  Rolston, B.~I. Schneider, and W.~D. Phillips, Science {\bf 287},  97  (2000).

\bibitem{burger1}
S.~Burger, K.~Bongs, S.~Dettmer, W.~Ertmer, K.~Sengstock, A.~Sanpera, G.~V. Shlyapnikov,
  and M.~Lewenstein, Phys. Rev. Lett. {\bf 83},  5198  (1999).

\bibitem{haus1}
H. Haus and W.~S. Wong, Rev. Mod. Phys. {\bf 68},  423  (1996).

\bibitem{dalfovo1}
F. Dalfovo, S. Giorgini, L.~P. Pitaevskii, and S. Stringari, Rev. Mod. Phys.
  {\bf 71},  463  (1999).

\bibitem{koehler1}
T. K\"ohler, J. Phys. B {\bf 34},  L534  (2001).

\bibitem{sulem1}
C. Sulem and P.~L. Sulem, {\em Nonlinear Schr\"odinger Equations: Self-focusing
  Instability and Wave Collapse} (Springer-Verlag, New York, 1999).

\bibitem{dodd1}
R.~J. Dodd, M.~Edwards, C.~J. Williams, C.~W. Clark, M.~J. Holland, P.~A.
  Ruprecht, and K.~Burnett, Phys. Rev. A {\bf 54},  661  (1996).

\bibitem{kagan2}
Y. Kagan, E.~L. Surkov, and G.~V. Shlyapnikov, Phys. Rev. Lett. {\bf 79},  2604
   (1997).

\bibitem{sackett2}
C.~A. Sackett, H.~T.~C. Stoof, and R.~G. Hulet, Phys. Rev. Lett. {\bf 80},
  2031  (1998).

\bibitem{bradley1}
C.~C. Bradley, C.~A. Sackett, J.~J. Tollett, and R.~G. Hulet,
\newblock {\em Phys. Rev. Lett.} {\bf 75}, 1687  (1995).

\bibitem{bradley2}
C.~C. Bradley, C.~A. Sackett, and R.~G. Hulet,
\newblock {\em Phys. Rev. A} {\bf 55}, 3951  (1997).

\bibitem{vogels1}
J.~M. Vogels, C.~C. Tsai, R.~S. Freeland, S.~J. J. M.~F. Kokkelmans, B.~J.
  Verhaar, and D.~J. Heinzen,
\newblock {\em Phys. Rev. A} {\bf 56}, R1067  (1997).

\bibitem{roberts1}
J.~L. Roberts, N.~R. Claussen, S.~L. Cornish, E.~A. Donley, E.~A. Cornell, and
  C.~E. Wieman, Phys. Rev. Lett. {\bf 86},  4211  (2001).

\bibitem{donley1}
E.~A. Donley, N.~R. Claussen, S.~L. Cornish, J.~L. Roberts, E.~A. Cornell, and
  C.~E. Wieman, e-print cond-mat/010519 (unpublished).

\bibitem{sackett1}
C.~A. Sackett, J.~M. Gerton, M. Welling, and R.~G. Hulet, Phys. Rev. Lett. {\bf
  82},  876  (1999).

\bibitem{carr29}
L.~Khaykovich, F.~Schreck, F.~Ferrari, T.~Bourdel, J.~Cubizolles, L.~D. Carr,
  Y.~Castin, and C.~Salomon, Science {\bf 296},  1290  (2002).

\bibitem{strecker1}
K.~E. Strecker, G.~B. Partridge, A.~G. Truscott, and R.~G. Hulet, Nature {\bf
  417},  150  (2002).

\bibitem{band1}
Y.~B. Band, B. Malomed, and M. Trippenbach, Phys. Rev. A {\bf 65},  033607
  (2002).

\bibitem{stamperkurn1}
D.~M. Stamper-Kurn {\it et~al.}, Phys. Rev. Lett. {\bf 83},  661  (1999).

\bibitem{desaix1}
M. Desaix, D. Anderson, and M. Lisak, J. Opt. Soc. Am. B {\bf 8},  2082
  (1991).

\bibitem{perez2}
V.~M. P\'erez-Garc\'ia, H.~Michinel, J.~I. Cirac, M.~Lewenstein, and P.~Zoller, Phys. Rev. A {\bf 56},  1424  (1997).

\bibitem{baym1}
G. Baym and C.~J. Pethick, Phys. Rev. Lett. {\bf 76},  6  (1996).

\bibitem{ruprecht1}
P.~A. Ruprecht, M.~J. Holland, K.~Burnett, and Mark Edwards,
\newblock {\em Phys. Rev. A} {\bf 51}, 4704 (1995).

\bibitem{eleftheriou1}
A.~Eleftheriou and K.~Huang,
\newblock {\em Phys. Rev. A} {\bf 61}, 043601 (2000).

\bibitem{castin2}
Y. Castin,  in {\em Coherent atomic matter waves}, edited by R. Kaiser, C.
  Westbrook, and F. David (EDP Sciences and Springer-Verlag, Les Ulis, France
  and Berlin, Germany, 2001), pp.\ 1--136, e-print cond-mat/0105058.

\bibitem{olshanii1}
M.~Olshanii,
\newblock {\em Phys. Rev. Letts.} {\bf 81}, 938 (1998).

\bibitem{press1}
W.~H. Press, S.~A. Teukolsky, W.~T. Vetterling, and B.~P. Flannery, {\em
  Numerical Recipes in C: The Art of Scientific Computing} (Cambridge Univ.
  Press, Cambridge, U.K., 1993).

\bibitem{note1}
Specifically, in one-dimensional simulations of Eq.~(\ref{eqn:gpe1d}),
an external potential of form $V_{{\rm abs}}=-iA\sin^2(\pi/2(
|x|-b)/(L/2-b))$ for $|x|>b$ and $V_{{\rm abs}}=0$ for $|x|<b$ is
added to a box of length $L$ centered at the origin.  For simulation parameters $\eta\simeq 1.0$, $L=64.0$, and
$\omega_z^2=-0.1$ the absorbing bound parameters $A=1.0$
and $b=(5/8)(L/2)$ were found to be sufficient.  In the case where
$\omega_z^2=0$, one must avoid the reflection of the low energy matter
waves induced by the spatial variation of $V_{{\rm abs}}$ by adding a
real potential drop far from the center of
the soliton and before the absorbing bounds; a sinusoidal function
squared was used.

\bibitem{singh1}
K.~G. Singh and D.~S. Rokhsar, Phys. Rev. Lett. {\bf 77},  1667  (1996).

\bibitem{satsuma1}
J. Satsuma and Y. N., Prog. of Theor. Phys. (Suppl.) {\bf 55},  284  (1974).

\bibitem{castin3}
Y. Castin and C. Herzog (unpublished).

\bibitem{kaup1}
D.~J. Kaup, Phys. Rev. A {\bf 42},  5689  (1990).

\bibitem{olshaniicommunication1}
1997, M. Olshanii, Univ. of Boston, private communication.

\bibitem{landau2}
L.~D. Landau and E.~M. Lifshitz, {\em Quantum Mechanics (Non-relativistic
  Theory)} (Pergamon Press, Tarrytown, New York, 1977), Vol.~3.

\bibitem{gordon1}
J.~P. Gordon, Opt. Lett. {\bf 8},  596  (1983).

\bibitem{carr19}
L.~D. Carr, J.~N. Kutz, and W.~P. Reinhardt, Phys. Rev. E {\bf 63},  066604
  (2001).

\end{thebibliography}

%\bibitem[*]{byline} to whom correspondence should be addressed

\end{document}